\documentclass[preprint]{elsarticle}

\usepackage{graphicx}

\begin{document}

\title{
Conjectured Exact Percolation Thresholds of the Fortuin-Kasteleyn Cluster for the $\pm J$ Ising Spin Glass Model
}

\author{
Chiaki \textsc{Yamaguchi}%
}

\address{
Kosugichou 1-359, Kawasaki 211-0063, Japan
}

\begin{abstract}
 The conjectured exact percolation thresholds of the Fortuin-Kasteleyn cluster 
 for the $\pm J$ Ising spin glass model
 are theoretically shown based on a conjecture.
 It is pointed out that
 the percolation transition of the Fortuin-Kasteleyn cluster 
 for the spin glass model
 is related to a dynamical transition for the freezing of spins.
 The present results are obtained as locations of points on the so-called Nishimori line,
 which is a special line in the phase diagram.
 We obtain
 $T_{FK} = 2 / \ln [z / (z - 2)]$ and $p_{_{FK}} = z / [2 (z - 1)]$
 for the Bethe lattice,
 $T_{FK} \to \infty$ and $p_{_{FK}} \to 1 / 2$
 for the infinite-range model,
 $T_{FK} = 2 / \ln 3$ and $p_{_{FK}} = 3 / 4$
 for the square lattice,
 $T_{FK} \sim 3.9347$ and $p_{_{FK}} \sim 0.62441$
 for the simple cubic lattice,
 $T_{FK} \sim 6.191$ and $p_{_{FK}} \sim 0.5801$
 for the 4-dimensional hypercubic lattice, 
 and
 $T_{FK} = 2 / \ln \{ [1 + 2 \sin (\pi / 18)] / [1 - 2 \sin (\pi / 18) ] \}$
 and $p_{_{FK}} = [1 + 2 \sin (\pi / 18) ] / 2$
 for the triangular lattice,
 when $J / k_B = 1$,
 where $z$ is the coordination number,
 $J$ is the strength of the exchange interaction between spins,
 $k_B$ is the Boltzmann constant,
 $T_{FK}$ is the temperature at the percolation transition point,
 and
 $p_{_{FK}}$ is the probability, that the interaction is ferromagnetic,
 at the percolation transition point.
\end{abstract}


\maketitle

\section{Introduction}

To establish reliable analytical theories of spin glasses has
 been one of the most challenging problems in statistical physics 
 for years\cite{N1, MH, N2, HM, O, EA, Y3}.
 Our main interest in this article
 does not lie directly in the issue of 
 the properties of the phases in spin glasses.
 We instead will
 concentrate ourselves on the precise
 determination of the structure
 of phase diagram for spin glasses. This problem
 is of practical importance 
 for numerical studies,
 since exact locations of 
 transition points greatly
 facilitate reliable estimates
 of physical properties
 around the transition points.

The $\pm J$ model is known
 as one of the Ising spin glass models\cite{N1, MH, N2, HM, O, Y3}.
In this article, the percolation transition of the Fortuin-Kasteleyn (FK) cluster is
investigated. The FK cluster has the FK representation\cite{FK, KF}.
In the ferromagnetic
Ising model, the percolation transition point agrees with the phase transition point\cite{CK}.
The $\pm J$ model has a
conflict in the interactions: the percolation transition point disagrees with the phase
transition point. However, it was pointed out by de
Arcangelis et al.\cite{ACP} that the correct understanding of the percolation phenomenon of
the FK cluster in the Ising spin glass model is important since a dynamical
transition occurs at a temperature very close to the percolation temperature, and
the dynamical transition and percolation transition are related to a transition for a
signal propagating between spins.
 The dynamical transition is a transition for the freezing of spins,
 which is investigated by a distance called
 the damage or the Hamming distance\cite{ACP, CB, DW, D}.
 
 A line in the phase diagram for the $\pm J$ model is called the Nishimori line\cite{N2}.
 The internal energy, the upper bound of the specific heat and so forth
 are exactly calculated on the line\cite{N1, MH, N2, HM, O, Y3}.
 In addition,
 the internal energy does not depend on any lattice shape 
 and instead depends on the number of nearest-neighbor pairs in the
 whole system.
 The location of the multicritical point can be on the Nishimori line, 
 is conjectured on the square lattice, and
 the conjectured location is in good agreement with
 the results of other numerical estimates\cite{NN}.
 The present results are also obtained as locations of points on the Nishimori line.

We use a conjecture and the values of the threshold fractions of
 the random bond percolation problem for obtaining results in this article.
 If a threshold fraction of
 the random bond percolation problem is calculated,
 one is able to calculate the conjectured percolation threshold
 of the FK cluster in the $\pm J$ model by using the present theory.
 Generally, calculation of the threshold fractions of
 the random bond percolation problem is easier than
 that of the percolation thresholds of the FK cluster in the $\pm J$ model.
 Therefore, the present theory can be promising in this respect.

This article is organized as follows.
 In \S\ref{sec:2}, the $\pm J$ model is explained.
 In \S\ref{sec:3},
 our conjecture is described, and conjectured exact equations are shown.
 The present results by using the obtained equations
 are given in \S\ref{sec:4}.
 In \S\ref{sec:5}, the concluding remarks of
 this article are described.

\section{Model} \label{sec:2}

The Hamiltonian for the $\pm J$ model, ${\cal H}$, 
 is given by\cite{N1, EA}
\begin{equation}
 {\cal H} = - \sum_{\langle i, j \rangle} J_{i, j} S_i S_j \, ,
\end{equation}
 where $\langle i, j \rangle$ denotes nearest-neighbor pairs, 
 $S_i$ is a state of the spin at site $i$, 
 and $S_i = \pm 1$.
 $J_{i, j}$ is the strength of the exchange interaction between
 the spins at sites $i$ and $j$.
 The value of $J_{i, j}$ is given with a distribution $P (J_{i, j})$.
 The distribution $P (J_{i, j})$
 is given by
\begin{equation}
 P (J_{i, j}) = p \, \delta_{J_{i, j}, J} + (1 - p) \, \delta_{J_{i, j}, - J}
 \, , \label{eq:PpmJJij}
\end{equation}
 where $J > 0$, and $\delta$ is the Kronecker delta.
 $p$ is the probability that the interaction
  is ferromagnetic, and $1-p$ is the probability
 that the interaction is antiferromagnetic.

We apply a percolation theory.
 We use the Fortuin-Kasteleyn (FK) cluster\cite{FK, KF}.
 The FK clusters consist of the FK bonds which are probabilistically put between spins.
 The number of the FK bonds is rigorously related to the internal energy\cite{Y3}.
 We define the probability for putting the FK bond
 as $P_{FK}$.
 $P_{FK}$ is given by\cite{ACP, Y}
\begin{equation}
 P_{FK}  = 1 -
 e^{- \beta J_{ij} S_i S_j - \beta |J_{ij}|} \, , \label{eq:PFKEAI}
\end{equation}
 where $\beta = 1 / k_B T$, $T$ is the temperature,
 and $k_B$ is the Boltzmann constant.
 For calculating $[\langle P_{FK} \rangle_T ]_R$, 
 a gauge transformation is used,
 where $\langle \, \rangle_T$ denotes the thermal average.
 The gauge transformation is performed by\cite{N1, T}
\begin{equation}
 J_{i, j} \to J_{i, j} \sigma_i \sigma_j \, , \quad S_i \to S_i \sigma_i \, ,
 \label{eq:GaugeT} 
\end{equation}
 where $\sigma_i$ is a variable at site $i$, and  $\sigma_i = \pm 1$.
 The gauge transformation has no effect
 on thermodynamic quantities\cite{T}.
 By performing the gauge transformation,
 the ${\cal H}$ part becomes ${\cal H} \to {\cal H}$ and
 the $P_{FK}$ part becomes $P_{FK} \to P_{FK}$.
 By using Eq.~(\ref{eq:PpmJJij}), the distribution $P (J_{i, j})$ 
 is rewritten as\cite{N1}
\begin{equation}
 P (J_{i, j}) = \frac{e^{\beta_P J_{i, j}}}
{2 \cosh (\beta_P J)} \, , 
 \quad J_{i, j} = \pm J \, , \label{eq:PpmJJij2}
\end{equation}
 where $\beta_P$ is given by
\begin{equation}
 \beta_P = \frac{1}{2 J} \ln \frac{p}{1-p} \, . \label{eq:betaPpmJ}
\end{equation}
 By performing the gauge transformation,  
 the distribution $P (J_{i, j})$ part
 becomes
\begin{eqnarray}
 \prod_{\langle i, j \rangle} P (J_{i, j})
 &=& \frac{e^{\beta_P \sum_{\langle i, j \rangle} J_{i, j}}}
{[2 \cosh (\beta_P J)]^{N_B}}  \nonumber \\
 &\to& \frac{\sum_{\{ \sigma_i \}} e^{\beta_P \sum_{\langle i, j \rangle}
 J_{i, j} \sigma_i \sigma_j}}
 {2^ N [2 \cosh (\beta_P J)]^{N_B}} \, ,
 \label{eq:PpmJJij3}
\end{eqnarray}
 where $N_B$ is the number of nearest-neighbor pairs in the
 whole system.
 When the value of $\beta_P$
 is consistent with the value of 
 the inverse temperature $\beta$, 
 the line for $\beta = \beta_P$ in the phase diagram
 is called the Nishimori line\cite{N2}.
 By using the gauge transformation,
 $[\langle P_{FK} \rangle_T ]_R$ on the Nishimori line 
 is obtained as\cite{Y3, Y}
\begin{equation}
 [\langle P_{FK} \rangle_T ]_R = \tanh (\beta_P J) \, , \label{eq:PFKpmJ}
\end{equation}
 where $\beta_P = 1 / k_B T_P$, and $T_P$ is the temperature on the Nishimori line.
 The value of $[\langle P_{FK} \rangle_T ]_R$ 
 on the Nishimori line does not depend on
 any lattice shape.

The dynamical transition mentioned in this article
 is characterized by
 a distance between two spin configurations $\{ S_i \}$ and $\{ \tilde{S}_i \}$.
 By using the distance, the freezing of spins is investigated.
 The distance $D (t)$ is given by\cite{DW, D}
\begin{equation}
 D (t) = \frac{1}{2 N}
 [ \langle \sum_i^{N} | S_i (t) - \tilde{S}_i (t) | \rangle_{\rm MC} ]_R \, ,
\end{equation}
 where $t$ is the time, $N$ is the number of sites,
 $\langle \, \rangle_{\rm MC}$ denotes the sample average
 by the Monte Carlo method, and
 $[ \, ]_R$ denotes the random configuration average
 for the exchange interactions.
 As for the initial condition of $\{ S_i \}$ and $\{ \tilde{S}_i \}$,
 $\{ S_i (0) \}$ is set to random,
 and $\{ \tilde{S}_i (0) \} = - \{ S_i (0) \}$ is set, for example.
 As for the initial condition, 
 the two configurations are generally 
 set to have a relationship with each other.
 As for the Monte Carlo method, a heat-bath method is used:
 by using the uniform pseudo-random number $r (t)$ $(0 \le r (t) < 1)$,
 update of $\{ S_i (t) \}$ and $\{ \tilde{S}_i (t) \}$ is given by
\begin{eqnarray}
 S_i (t + \Delta t ) &=& {\rm sign} \biggl\{ 
 \frac{1}{1 + \exp [- 2 \beta \sum_{j} J_{i, j} S_j (t) ]}
 - r (t) \biggr\} \, , \label{eq:Si1} \\
 \tilde{S}_i (t + \Delta t ) &=& 
 {\rm sign} \biggl\{ 
 \frac{1}{1 + \exp [- 2 \beta \sum_{j} J_{i, j} \tilde{S}_j (t) ]}
 - r (t) \biggr\} \, , \label{eq:Si2} 
\end{eqnarray}
 where $\Delta t$ is a time step per spin,
 the summations of the right-hand sides of Eqs.~(\ref{eq:Si1}) and (\ref{eq:Si2})
 are over the nearest-neighbor sites of the site $i$.
 The two spin configurations $\{ S_i (t) \}$ and $\{ \tilde{S}_i (t) \}$
 are updated with the same pseudo-random number sequence $\{ r (t) \}$.
 It has been pointed out that, for the cubic lattice, 
 there are three phases\cite{DW}, i.e., a high-temperature phase,
 an intermediate phase and a low-temperature phase.
 In the high-temperature phase,
 the two configurations become identical quickly, so that
 the distance between them vanishes.
 In the intermediate phase and the low-temperature phase,
 the distance between the two configurations remains a finite
 in the long-time limit if the system size is large enough.
 In the intermediate phase, the distance between the two configurations
 does not depend on the initial conditions of the two configurations.
 In the low-temperature phase, the distance between the two configurations
 depends on the initial conditions of the two configurations.
 It is pointed out that the dynamical transition temperature $T_D$
 between the high-temperature phase and the intermediate phase is
 related to the percolation transition temperature $T_{FK}$
 of the FK cluster\cite{ACP, CB}.
 It is pointed out in Refs.\cite{ACP, CB} that there is a possibility that
 the value of  $T_{FK}$ is consistent with the value of $T_D$.
 Recently, it was pointed out in Ref.\cite{LC} that, by a Monte Carlo calculation,
 the value of  $T_{FK}$ is not consistent with the value of $T_D$, and
 the value of $T_{FK}$ is an approximated value of $T_D$.

\section{Conjecture} \label{sec:3}

We use a conjecture.
 If the conjecture is correct,
 the present theory gives the exact values.
 Here, we describe the conjecture
 and show conjectured exact equations derived by using the conjecture.

 We conjecture
\begin{equation}
 [\langle P_{FK} \rangle_T ]_R =
 P_C  \label{eq:conjecture1}
\end{equation}
 at the percolation transition point of the FK cluster
 on the Nishimori line for arbitrary lattices,
 where $P_{FK}$ is the probability for putting the FK bond between spins,
 and $P_C$ is the threshold fraction of the random bond percolation problem.  
 In the random bond percolation problem,
 bonds for generating clusters are randomly put on the edges of the lattice, and
 one of the clusters is percolated at the threshold fraction $P_C$ \cite{K}.
 From Eq.~(\ref{eq:PFKpmJ}),
 the value of $[\langle P_{FK} \rangle_T ]_R$ 
 on the Nishimori line does not depend on
 any lattice shape. From the fact,
 Eq.~(\ref{eq:conjecture1}) is conjectured.
 We propose this conjecture in this article
 as a conjecture which may be exact.

By using Eqs.~(\ref{eq:betaPpmJ}), (\ref{eq:PFKpmJ}) and
 (\ref{eq:conjecture1}),
 we obtain
\begin{eqnarray}
 T_{FK} &=& \frac{2 J}{k_B \ln [ (1 + P_C ) / (1 - P_C ) ]} \, , \label{eq:TFK-conjecture} \\ 
 p_{_{FK}} &=& \frac{1}{2} (1 + P_C ) \, . \label{eq:p-conjecture}
\end{eqnarray}
 Eqs.~(\ref{eq:TFK-conjecture}) and (\ref{eq:p-conjecture})
 are conjectured exact equations.
 By using Eqs.~(\ref{eq:TFK-conjecture}),
 (\ref{eq:p-conjecture}) and the value of the 
 threshold fraction $P_C$ of the random bond percolation problem,
 the values of the percolation transition temperature $T_{FK}$ and
 the percolation transition probability $p_{_{FK}}$ are calculated
 as the location of a point on the Nishimori line.
 The obtained values are conjectured exact values.
 Note that the percolation transition probability $p_{_{FK}}$
 is the probability that the interaction
 is ferromagnetic at the percolation transition point.

Campbell and Bernardi have derived an equation 
 for the energy $E$ in the $\pm J$ model and the threshold fraction $P_C$
 of the random bond percolation problem, i.e.,
 $E = J N_B \{ 1 - 2 P_C / [1 - \exp (- 2 \beta J)]  \}$
 on the assumption of a random active-bond spatial distribution\cite{CB}.
 By applying the energy and the temperature on the Nishimori line
 to this equation, the same equations 
 (Eqs.~(\ref{eq:TFK-conjecture}) and (\ref{eq:p-conjecture}))
 are obtained\cite{YC},
 where the energy on the Nishimori line is 
 $- N_B J \tanh (\beta_P J)$ \cite{N1}.

In Ref.\cite{Y},
 the exact values of the percolation threshold of the FK cluster
 are also conjectured.
 The conjecture proposed in this article is more general,
 so that
 the present study includes
 finite-dimensional cases that
 are not mentioned in Ref.\cite{Y}.

\section{Results} \label{sec:4}

We show the present results by applying
 the conjectured exact equations obtained in \S\ref{sec:3}, i.e.,
 Eqs.~(\ref{eq:TFK-conjecture}) and (\ref{eq:p-conjecture}),
 to several lattices.
 The present results are obtained as locations of points on the Nishimori line.

For the Bethe lattice,
 the threshold fraction $P_C$ 
 of the random bond percolation problem
 is obtained as $P_C = 1 / (z - 1)$ \cite{SA},
 where $z$ is the coordination number.
 By using Eqs.~(\ref{eq:TFK-conjecture}) and (\ref{eq:p-conjecture}),
 we obtain
$$
 T_{FK} =
 \frac{2 J}{k_B \ln [z / (z - 2)]} \, , \quad p_{_{FK}} = \frac{z}{2 (z - 1)} \, .
$$
 This is the result for the Bethe lattice.
 $T_{FK}$ agrees with the ferromagnetic transition temperature 
 for the pure system, $T_C$ \cite{B}.
 When $z = N - 1$ and  $J \to J / \sqrt{N}$,
 this model becomes the infinite-range model.
 Then, in the thermodynamic limit, 
 we obtain $T_{FK} \to \infty$ and $p_{_{FK}} \to 1 / 2$
 when $J / k_B = 1$.
 This result for the infinite-range model agrees with $T_{FK}$ of the
 previous result in Ref.\cite{MNS} and
 $T_D$ of the previous results in Refs.\cite{D, CA}.

For the square lattice,
 the threshold fraction of the random bond percolation problem
 is obtained as $1 / 2$ \cite{K}.
 By using Eqs.~(\ref{eq:TFK-conjecture}) and (\ref{eq:p-conjecture}),
 we obtain
 $$T_{FK} = \frac{2}{\ln 3} \sim 1.820478 \, , \quad p_{_{FK}} = \frac{3}{4} = 0.75 \, ,$$
 when $J / k_B = 1$.
 This is the result for the square lattice.
 When $J / k_B = 1$, the ferromagnetic transition temperature 
 for the pure system, $T_C$,
 is $2 / \ln (1 + \sqrt{2})$ ($ \sim 2.269$) \cite{B} .
 $T_{FK}$ does not agree with $T_C$ in this case.
 The previous numerical results are
 $T_{FK} = 1.81(2)$ and $1.82(3)$ \cite{IIK} for $p = 0.7$,
 $T_{FK} = 1.83(2)$ and $1.83(3)$ \cite{IIK} for $p = 0.8$,
 $T_D \sim 1.8$ \cite{D} for $p = 1 / 2$,
 $T_D \sim 1.70$ \cite{CA} for $p = 1 / 2$, and
 $T_D = 1.69(2)$ \cite{LC} for $p = 1 / 2$.
 This result agrees with
 the previous results for $T_{FK}$ in Ref.\cite{IIK}.
 This result does not contradict with the previous result for $T_D$
 in Ref.\cite{D}. However,
 it seems that this result does not agree
 with the previous results for $T_D$ in Refs.\cite{CA, LC}.

\begin{figure}[t]
\begin{center}
\includegraphics[width=0.60\linewidth]{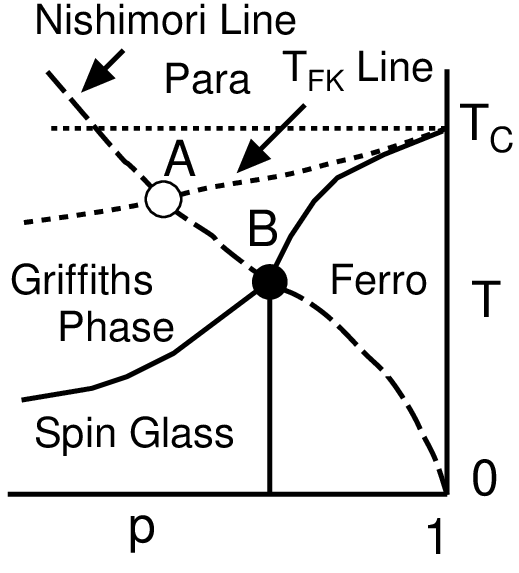}
\end{center}
\caption{
 A schematic phase diagram for the $\pm J$ model.
 $p$ is the probability that the interaction
 is ferromagnetic, and $1-p$ is the probability
 that the interaction is antiferromagnetic.
 $T$ is the temperature.
 The paramagnetic phase (`Para'),
 the ferromagnetic phase (`Ferro'), the Griffiths phase (`Griffiths Phase')
 and the spin glass phase (`Spin Glass') are depicted.
 The Nishimori line (the dashed line),
 the Griffiths temperature (the dotted line) 
 and the $T_{FK}$ line (the short dashed line)
 are also depicted. 
 $T_C$ is the Curie temperature for the ferromagnetic model.
 The point `A' is the percolation transition point
 of the FK cluster on the Nishimori line.
 The point `B' is the multicritical point.
\label{fig:phase-diagram}
}
\end{figure}
 Fig.\ref{fig:phase-diagram} shows
 a schematic phase diagram for the $\pm J$ model.
 $p$ is the probability that the interaction
 is ferromagnetic, and $1-p$ is the probability
 that the interaction is antiferromagnetic.
 $T$ is the temperature.
 The paramagnetic phase (`Para'), the ferromagnetic phase (`Ferro'),
 the Griffiths phase (`Griffiths Phase') and the spin glass phase (`Spin Glass')
 are depicted.
 In this article,
 we do not mention the existence of
 a mixed phase between the ferromagnetic phase
 and the spin glass phase.
 The Nishimori line (the dashed line),
 the Griffiths temperature (the dotted line)
 and the $T_{FK}$ line (the short dashed line)
 are also depicted. 
 $T_C$ is the Curie
 temperature for the ferromagnetic model.
 The value of the Griffiths temperature 
 corresponds to the value of $T_C$.
 The $T_{FK}$ line represents the percolation transition temperature of the FK cluster.
 $T_{FK}$ agrees with $T_C$ \cite{CK} when $p = 1$, and
 $T_D$
 numerically agrees very well with $T_C$ \cite{DW} when $p = 1$.
 The phase between the Griffiths temperature
 and the $T_{FK}$ line can also be called
 the Griffiths phase, but, it can be considered that the behavior of the distance in this phase
 is the same as that in the paramagnetic phase since $T_{FK} \sim T_D$.  
 The point `A' is the percolation transition point
 of the FK cluster on the Nishimori line.
 The point `B' is the multicritical point.
 In a certain lattice case, $T_{FK}$ does not agree with $T_C$. 
 On the other hand, in a certain lattice case,
 $T_{FK}$ agrees with $T_C$. 
 For example, 
 $T_{FK}$ for the square lattice does not agree 
 with $T_C$ for the same lattice as mentioned above.
 On the other hand, 
 $T_{FK}$ for the Bethe lattice 
 agrees with $T_C$ for the same lattice as mentioned above.
 For the multicritical point,
 it seems that the temperature at the multicritical point
 generally does not agree with $T_{FK}$.
 For example,
 when $J / k_B = 1$,
 the temperature at the multicritical point
 for the square lattice
 is roughly equal to $0.957$ \cite{NN},
 while $T_{FK}$ for the square lattice is
 $2 / \ln 3$ ( $\sim 1.82$) from the present result.
 In addition, the temperature at the multicritical point
 for the Bethe lattice
 is  $2 J / k_B \ln [(\sqrt{z -1} + 1) / (\sqrt{z - 1} - 1)]$
 from the result in Ref.\cite{CCCST},
 while $T_{FK}$ for the Bethe lattice is $2 J / k_B \ln [z / (z - 2)]$ from the present result.

For the simple cubic lattice,
 the threshold fraction of the random bond percolation problem is
 numerically estimated as $P_C \sim 0.248813$ \cite{LZ}.
 By using Eqs.~(\ref{eq:TFK-conjecture}) and (\ref{eq:p-conjecture}),
 we obtain
$$T_{FK} \sim 3.9347 \, , \quad p_{_{FK}} \sim 0.62441 \, ,$$
 when $J / k_B = 1$.
 This is the result for the simple cubic lattice.
 The previous numerical results 
 are $T_{FK} \sim 3.92$ \cite{ACP} for $p = 1 / 2$,
 $T_D \sim 4.1$ \cite{DW} for $p = 1 / 2$,
 $T_D \sim 4.0$ \cite{D} for $p = 1 / 2$,
 $T_D \sim 3.92$ \cite{CA} for $p = 1 / 2$, and
 $T_D = 3.932(2)$ \cite{LC} for $p = 0.6244$.
 This result does not contradict with the previous results for $T_{FK}$ and $T_D$
 in Refs.\cite{ACP, CA}. This result slightly disagree with
 the previous result for $T_D$ in Ref.\cite{LC}, but it seems that
 the difference between the two results is too small to judge the consistency.
 It seems that this result does not agree
 with the previous results for $T_D$ in Refs.\cite{DW, D}.

For the 4-dimensional hypercubic lattice,
 the threshold fraction of the random bond percolation problem is
 numerically estimated as $P_C \sim 0.16013$ \cite{PZS}.
 By using Eqs.~(\ref{eq:TFK-conjecture}) and (\ref{eq:p-conjecture}),
 we obtain
$$T_{FK} \sim 6.191 \, , \quad p_{_{FK}} \sim 0.5801 \, ,$$
 when $J / k_B = 1$.
 This is the result for the 4-dimensional hypercubic lattice.
 The previous numerical results 
 are $T_D \sim 6$ \cite{D} for $p = 1 / 2$,
 $T_D \sim 6.03$ \cite{CA} for $p = 1 / 2$, and
 $T_D = 6.057(10)$ \cite{LC} for $p = 0.58006$.
 This result does not contradict with the previous results for $T_D$
 in Refs.\cite{D, CA}, and this result slightly disagree with the previous result for $T_D$
 in Ref.\cite{LC}.

For the triangular lattice,
 the threshold fraction of the random bond percolation problem
 is obtained as $2 \sin (\pi / 18)$ \cite{SE}.
 By using Eqs.~(\ref{eq:TFK-conjecture}) and (\ref{eq:p-conjecture}),
 we obtain
 $$T_{FK} = \frac{2 }
{ \ln \bigl[ \frac{1 + 2 \sin (\pi / 18) }{ 1 - 2 \sin (\pi / 18)} \bigr] }
 \sim 2.759641
  \, , \quad p_{_{FK}} = \frac{1 + 2 \sin (\pi / 18) }{ 2 } \sim 0.6736482 \, ,$$
 when $J / k_B = 1$.
 This is the result for the triangular lattice.
 The previous numerical results 
 are $T_{FK} = 2.74(3)$ and $2.73(4)$ \cite{IIK} for $p = 0.6$, and
 $T_{FK} = 2.75(3)$ and $2.76(3)$ \cite{IIK} for $p = 0.7$.
 This result agrees with
 the previous results for $T_{FK}$ in Ref.\cite{IIK}.

The present result is possibly exact for $T_{FK}$.

\section{Concluding Remarks} \label{sec:5}

We theoretically showed 
 the conjectured exact percolation thresholds of the FK cluster 
 for the $\pm J$ Ising spin glass model based on a conjecture.
 The present theory possibly gives the exact values of
 the percolation transition temperature $T_{FK}$ on the Nishimori line.
 If the present theory is correct, one can obtain the possibly exact values \cite{ACP, CB}
 or approximated values \cite{LC}
 of the dynamical transition temperature $T_D$ for the freezing of spins on the Nishimori line
 by using $T_{FK} \sim T_D$.

The present theory may be applied directly to the Gaussian Ising spin glass model and
 the Potts gauge glass model.
 Concretely,
 the solution of $[ \langle P_{FK} \rangle_T ]_R$ on the Nishimori line
 for the Gaussian Ising spin glass model is obtained in Ref.\cite{Y},
 and the solution of $[ \langle P_{FK} \rangle_T ]_R$ on the Nishimori line 
 for the Potts gauge glass model is obtained in Ref.\cite{Y2}.
 By using these solutions instead of the solution for the $\pm J$ Ising spin glass model,
 the conjectured exact equations in the Gaussian Ising spin glass model and
 the Potts gauge glass model
 are obtained.

Recently, it was numerically shown in Ref.\cite{LC} that
 $T_{FK} / T_D$ on the Nishimori line for the simple cubic lattice,
 a four-dimensional lattice and a five-dimensional lattice
 are $1.0005(5)$, $1.022(2)$ and $1.036(10)$ respectively.
 So, it was concluded that $T_{FK}$ does not equal to $T_D$
 and the value of $T_{FK}$ is an approximated value of $T_D$.

We also mentioned the dynamical transition for the freezing of spins,
 which is investigated by the time evolution
 of the distance between two spin configurations on the Nishimori line.
 This study is different from the study of
 the aging phenomena on the Nishimori line
 as in Ref.\cite{O}.

\section*{Acknowledgments}

The author would like to thank I. Campbell for useful comments.

\end{document}